\documentclass{singlecol-new}

\usepackage{natbib,stfloats}
\usepackage{mathrsfs}

\usepackage{amsmath}
\usepackage{algorithm}
\usepackage{algorithmic}
\usepackage[utf8]{inputenc}
\usepackage{todonotes}
\usepackage{colortbl}
\usepackage{graphics}
\usepackage{epstopdf}
\usepackage{verbatim}
\usepackage{rotating}
\usepackage[inline]{enumitem}
\usepackage{color}
\usepackage{soul}

\makeatletter
\def\theequation{\arabic{equation}}

\makeatother

\begin{document}%

\setcounter{page}{1}

\LRH{F.F.B. Ara\'ujo, A.M. Costa, C. Miralles}

\RRH{Balancing parallel assembly lines with disabled workers}

\VOL{x}

\ISSUE{x}

\PUBYEAR{xxxx}

\BottomCatch


\PUBYEAR{2014}

\subtitle{}

\title{Balancing parallel assembly lines with disabled workers}

\authorA{Felipe F. B. Ara\'ujo*}
\affA{Instituto de Ci\^encias Matem\'aticas e da Computa\c{c}\~ao,\\ Universidade de S\~ao Paulo,\\ S\~ao Carlos, SP, Brazil \\
E-mail: felipefba@hotmail.com\\
*Corresponding author}
\authorB{Alysson M. Costa}
\affB{Department of Mathematics and Statistics,\\ University of Melbourne,\\ Melbourne, VIC, Australia \\
E-mail: alysson.costa@unimelb.edu.au}
\authorC{Crist\'obal Miralles}
\affC{ROGLE - Dpto. Organizaci\'{o}n de Empresas,\\ Universitat Polit\`ecnica de Val\`encia,\\ Valencia, Spain\\
E-mail: cmiralles@omp.upv.es}


\begin{abstract}
In this paper we study an assembly line balancing problem that occurs in sheltered worker centers for the disabled, where workers with very different characteristics are present. We are interested in the situation in which complete parallel assembly lines are allowed and name the resulting problem as parallel assembly line worker assignment and balancing problem (PALWABP). This approach enables many new possible worker-tasks assignments, what is beneficial in terms of both labour integration and productivity. We present a linear mixed-integer formulation and two heuristic solution methods: one is based on tabu search and the other is a biased random-key genetic algorithm (BRKGA). Computational results with a large set of instances recently proposed in the literature show the advantages of allowing such alternative line layouts. 
\end{abstract}

\KEYWORD{parallel assembly line balancing; heterogeneous workers; heuristics}

\REF{to this paper should be made as follows: Ara\'ujo, F.F.B. et al. (xxxx) `Balancing parallel assembly lines with disabled workers', {\it European Journal of Industrial Engineering}, Vol. x, No. x, pp.xxx\textendash xxx.}

\begin{bio}
Felipe F. B. Ara\'ujo received a MSc in computer science and computational mathematics from Universidade de S\~ao Paulo in 2011. He is currently a PhD student whose main interests are applications of operations research.\vs{9}

\noindent Alysson M. Costa received a PhD in Business Administration from HEC Montreal - University of Montreal in 2006. Currently, he is a Lecturer at the Department of Mathematics and Statistics - University of Melbourne. His research interests are mainly in applications of operations research.\vs{8}

\noindent Crist\'obal Miralles received his PhD in Industrial Engineering from Universitat Polit\`ecnica de Val\`encia, where he is lecturer at the Industrial Engineering faculty. He is member of ROGLE Research Group and his main interests are related with social/environmental applications of OR/MS, widening the classical profit-oriented approaches.
\end{bio}

\maketitle

\section{Introduction}\label{introduction}

According to the International Labour Organization (ILO), people with disabilities represent an estimated 10 per cent of the world's population, where more than 500 million are of working age. Labour market inclusion is easier to address in periods of increasing labour demand than in times of recession. Nevertheless, even in face of recent crisis period there is evidence of national actions to create more flexible work solutions. In this sense, one of the actions most commonly adopted to facilitate the integration of people with disabilities into the labor market has been the creation of Sheltered Work centers for Disabled (henceforth SWDs).

This model of socio-labor integration tries to move away from the traditional stereotype that considers people with disabilities unable to develop continuous professional work \citep{miralles10operations}. In countries such as Spain, for example, this labor integration formula keeps being successful in offering jobs to disabled people, and a common strategy of SWDs to facilitate this integration is the use of assembly lines as the most accessible configuration. In this sense \cite{miralles07advantages} were the first to evidence how the division of work in single tasks enables many possible job assignments that can make the disabilities invisible, even becoming a good method for therapeutic rehabilitation if appropriate job rotation mechanisms are applied \citep{costa09job}.

\subsection{Literature review and motivation of this work}

Traditional assembly line balancing research has focused on the simple assembly line balancing problem (SALBP), that uses several well-known simplifying hypotheses, which reduce the complex problem of assembly line configuration to the ``core'' problem of assigning tasks to stations so that certain precedence constraints are fulfilled. Thus, in the last decade a big effort has been made towards modeling real world assembly line systems through different extensions of SALBP, aiming to narrow the former gap between research and practice (see reviews of \cite{scholl06state}, \cite{becker06survey}, \cite{boysen07classification,boysen08assembly} or more recently \cite{dolgui13taxonomy}).

In this sense, the so-called assembly line worker assignment and balancing problem (ALWABP) represents one of these recent efforts made by the Academia. This approach focus on the heterogeneity of task times and the presence of incompatibilities, defining a new set of realistic hypotheses inspired by the SWD assembly lines where disabled workers execute tasks at different rates \citep{miralles07advantages}. Thus, the ALWABP defines worker-dependent processing times, which allow taking into account the diversity of workers and can, therefore, be useful in environments other than SWDs where workers also have diverse speeds to perform certain tasks \citep{araujo12two}.

Since the original paper of \cite{miralles07advantages}, many other references have contributed to give this problem visibility in the scientific literature, and several solution methods have been developed. Exact methods have focused on branch-and-bound strategies \citep{miralles08branch,borba14heuristic, vila14branch} although most of the efforts have concentrated on heuristic and metaheuristic methods due to the NP-hard condition of the problem, using strategies such as clustering search \citep{chaves09hybrid}, tabu search \citep{moreira09minimalist}, Iterated Beam Search \citep{blum11solving}, constructive heuristics \citep{moreira12simple} or genetic algorithms \citep{moreira12hybrid,mutlu13iterative}.

However, this intense research on the ALWABP is traditionally limited to the hypothesis of a single serial assembly line, whereas more variants would help SWD managers to handle the workforce specific requirements while optimizing productive efficiency. It has to be noted that SWDs compete in real markets and then:
\begin{enumerate*}[label=\itshape\alph*\upshape)]
	\item must be as efficient as possible to adapt to market fluctuations or simply to survive;
	\item receive some governmental help to compensate the lower yields of its workforce (with minimum of 70\% being disabled) and where certain limitations and therapy requirements must be considered.
\end{enumerate*}
Hence, this double aim makes it important for the SWD manager to count on as many alternative job assignments as possible.

In this context, this paper explores the possibility of assigning workers to teams, which can then operate at parallel assembly lines. In fact, by exploring this new configuration several advantages are obtained:

\begin{itemize}
	\item Higher global productivity levels shall be obtained, which is crucial for the survival of a SWD (or its growing, thus promoting new jobs for more people with disabilities): indeed, the possibility of designing lines in parallel increases the number of different available feasible solutions, introducing potentially many new assignment possibilities due to the combinatorial characteristic of the problem and circumventing limitations related to the combination of precedence constraints, worker x tasks incompatibilities and cycle time limits. Our hypothesis is that at least one of these solutions has a better assignment of tasks to workers that can reduce the total task execution time and/or reduce workstation idle times, increasing total productivity. 

	\item With parallel lines most workers will probably be assigned to a larger number of tasks, resulting in longer work schedules. It is also possible to assign more than one worker to the same task (in a different parallel line) in the same period: as it is stated in (Miralles et al. 2007) the primary SWD aim is to promote a work environment that helps disabled workers to have a positive and constant evolution in their own capabilities, in order to integrate them as soon as possible in ordinary Work Centers". Therefore, these more complex work schedules will certainly contribute to this evolution, empowering the workers and also avoiding too specialized work routines (that often provoke mental stress, especially in certain physical rehabilitation cases).
	\item Moreover, in the case of workers with very slow operation times, now they can be easily integrated with low productivity loses: as it happens with SALBP, when we allow parallelization the cycle time can be lower than the largest operation time, or in the case of ALWABP lower than the slowest task to execute (obtained with a max-min operation over all execution times).
	\item Finally, parallel lines provide an extra solution space for the manager, that now has more game to play when applying job rotation. Although the analysis of this tradeoff is out of the scope of this paper, some preliminary further research suggest that the effect of longer work routines at each station (which can eventually avoid certain rotations due to incompatibilities) is compensated by the very high number of new possible worker-tasks assignments arising. Thus, developing a decision support system that enlarge the classical job rotation approaches (see \cite{costa09job} or \cite{moreira12hybrid} becomes a promising future research line.
\end{itemize}

In front of these advantages, it must be noted that parallelization suppose the duplication of facilities and tools that may be necessary for different workers. This practical implication should be taken into account in those cases with physical constraints or very high installation costs; what is not usual for most SWDs lines, that typically get contracts to assemble low-added value products. 

The social dimension of SWDs further justify an analysis such as the one proposed here, where the cost of the line is relegated to a lower position. As certain social impacts are not easy to quantify, incurring in a precise estimation of the human resources cost would add little value to this study. Somehow it can be said that duplication costs, even when considerable, shall be compensated in the long term by the higher productivity reached and by the other qualitative advantages exposed above.

\subsection{Contribution and outline}

This last comment is also valid for \cite{araujo12two} that are, to the best of our knowledge, the sole authors to analyze parallelization in ALWABP. They study the assignment of workers in parallel or collaborating, but just at the same station of a single line. Although some positive results have been obtained, this approach has the drawback of increasing the complexity of coordination and control along (and within, in the case of collaboration) the stations. Parallelizing only certain stations and/or tasks can be cluttering because the production flow is divided/duplicated (what can be hard to coordinate, especially in the case of mental disabilities).

In this new approach, we propose parallelizing complete lines instead of single stations and/or tasks, making the assembly line management and self-control easier: every operator is assigned to a work team in charge of certain station within a complete parallel serial line with independent feeders, work in progress, buffers and tools. Thus, the (parallel) assembly lines allowed will make use of all available workers grouped by teams, where the additional decision on the teams to conform and the number of parallel lines will affect heavily the problem and, as will be later explained, the formal means to solve it properly. 

This new extension of ALWABP is named as the parallel assembly line worker assignment and balancing problem (PALWABP) and is presented in the next two sections: first by means of an example, and then with the formal definition and a mathematical model. Then in section \ref{methodologies} we present two heuristics for the problem: one based in Tabu Search (Section \ref{heuristics}) and the other an implementation of a Biased Random-Key Genetic Algorithm (Section \ref{brkga}). Computational results over a large set of instances are presented and analyzed in Section \ref{computational}. General conclusions and hints for future works end this paper in Section \ref{conclusions}.

\section{Parallel Assembly line Worker Assignment and Balancing Problem}

Most previous proposals in the literature face the ALWABP of type 2 (minimization of cycle time given a fixed number of workers), the most typical situation in reality, and have been mostly evaluated with the set of 320 benchmark instances first proposed by \cite{chaves09hybrid}. We have extracted the problem HESKIA 64 of this benchmark and expressed its input data in Table 1, where  for every task (rows) several operation times are possible depending on the worker (columns). If a task is considered unfeasible for certain worker, the incompatibility is represented by a dash in the corresponding cell of the input data matrix.

In order to show the benefits of the extension proposed here, its optimal solution is represented in Figure \ref{fig:ex_heskia_64}. Each arrow represents a precedence constraint. For example, the arrow pointing from task 1 to task 8 means that task 1 must be executed before task 8. We can notice how all the precedence constraints are respected while ensuring an optimal cycle time of 126 s (given by the bottleneck station, which is the second station with worker W6 performing task 20): 

\begin{figure}[!ht]
\centering
\resizebox{\textwidth}{!}{
\includegraphics{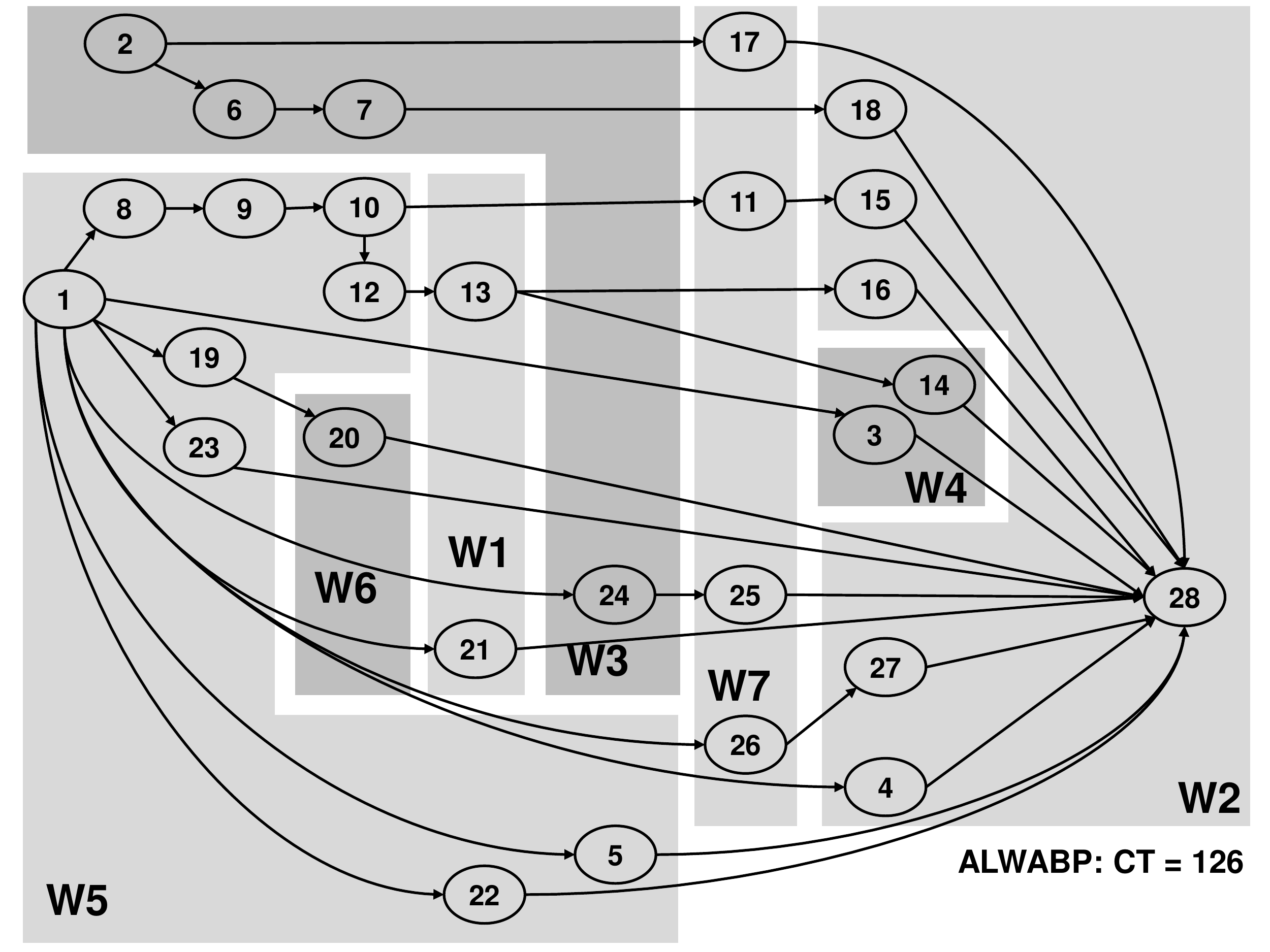}
}
\caption{Optimal solution for HESKIA\_64 with the traditional ALWABP approach}
\label{fig:ex_heskia_64}
\end{figure}

This is the minimum cycle time we can get with the traditional serial approach. But if we allow the possibility of designing complete parallel lines with the available workers, then solution space is enlarged, and lower combined cycle times can be reached. To illustrate this fact, let's focus now on table \ref{tab:heskia_64} (but now taking into account the light shaded/dark shaded cells, and the bottom of the table):

\begin{table}[!ht]
\centering
\resizebox{\textwidth}{!}{
\begin{tabular}{cccccccc}
\hline
{\bf Tasks:} &   {\bf W1} &   {\bf W2} &   {\bf W3} &   {\bf W4} &   {\bf W5} &   {\bf W6} &   {\bf W7} \\
\hline
   {\bf 1} &         70 &         88 &   \cellcolor[gray]{.6} 20 &         57 &          \cellcolor[gray]{.8} 5 &         42 &        118 \\
\hline
   {\bf 2} &         59 &         14 &   \cellcolor[gray]{.6} 17 &          - &         \cellcolor[gray]{.8} 17 &          2 &         20 \\
\hline
   {\bf 3} &         33 &         45 &   \cellcolor[gray]{.6} 26 &         13 &         \cellcolor[gray]{.8} 28 &         17 &          - \\
\hline
   {\bf 4} &          6 &          1 &          9 &          2 &          9 &    \cellcolor[gray]{.6} 1 &          \cellcolor[gray]{.8} 3 \\
\hline
   {\bf 5} &          1 &          2 &    \cellcolor[gray]{.6} 1 &          2 &          1 &          1 &          \cellcolor[gray]{.8} 1 \\
\hline
   {\bf 6} &         \cellcolor[gray]{.8} 27 &         25 &   \cellcolor[gray]{.6} 14 &         38 &         48 &         45 &         49 \\
\hline
   {\bf 7} &         17 &         20 &    \cellcolor[gray]{.6} 9 &          9 &         27 &         17 &         \cellcolor[gray]{.8} 32 \\
\hline
   {\bf 8} &         62 &         43 &   \cellcolor[gray]{.6} 97 &          9 &        \cellcolor[gray]{.8} 10 &        100 &         26 \\
\hline
   {\bf 9} &         31 &         56 &   \cellcolor[gray]{.6} 60 &          - &        \cellcolor[gray]{.8} 11 &          9 &          - \\
\hline
  {\bf 10} &         53 &         91 &   \cellcolor[gray]{.6} 36 &         37 &        \cellcolor[gray]{.8} 12 &        101 &         93 \\
\hline
  {\bf 11} &         21 &          - &   \cellcolor[gray]{.6} 13 &         17 &          7 &         35 &         \cellcolor[gray]{.8} 5 \\
\hline
  {\bf 12} &         19 &         31 &    \cellcolor[gray]{.6} 9 &         \cellcolor[gray]{.8} 1 &         15 &          1 &         36 \\
\hline
  {\bf 13} &       \cellcolor[gray]{.8} 108 &  \cellcolor[gray]{.6} 136 &          - &        194 &        133 &        179 &        171 \\
\hline
  {\bf 14} &         52 &   \cellcolor[gray]{.6} 76 &         32 &         68 &          - &          - &        \cellcolor[gray]{.8} 33 \\
\hline
  {\bf 15} &          5 &         10 &    \cellcolor[gray]{.6} 6 &          6 &          9 &          8 &         \cellcolor[gray]{.8} 4 \\
\hline
  {\bf 16} &          8 &          4 &          - &          7 &          2 &    \cellcolor[gray]{.6} 8 &         \cellcolor[gray]{.8} 2 \\
\hline
  {\bf 17} &         97 &          - &        165 &         99 &         \cellcolor[gray]{.8} 5 &  \cellcolor[gray]{.6} 150 &        103 \\
\hline
  {\bf 18} &          8 &          8 &   \cellcolor[gray]{.6} 12 &          1 &          - &         12 &         \cellcolor[gray]{.8} 4 \\
\hline
  {\bf 19} &         47 &         50 &   \cellcolor[gray]{.6} 22 &         \cellcolor[gray]{.8} 2 &         49 &          - &         25 \\
\hline
  {\bf 20} &         67 &  \cellcolor[gray]{.6} 126 &        129 &       \cellcolor[gray]{.8} 132 &          - &        126 &        133 \\
\hline
  {\bf 21} &         17 &          5 &         26 &         12 &         10 &   \cellcolor[gray]{.6} 15 &         \cellcolor[gray]{.8} 4 \\
\hline
  {\bf 22} &          8 &   \cellcolor[gray]{.6} 16 &          - &          6 &         15 &          - &        \cellcolor[gray]{.8} 12 \\
\hline
  {\bf 23} &          3 &          - &    \cellcolor[gray]{.6} 3 &          3 &         \cellcolor[gray]{.8} 4 &          - &          - \\
\hline
  {\bf 24} &         21 &         29 &         37 &         18 &         32 &   \cellcolor[gray]{.6} 12 &        \cellcolor[gray]{.8} 32 \\
\hline
  {\bf 25} &        107 &        163 &          - &        177 &        179 &   \cellcolor[gray]{.6} 94 &        \cellcolor[gray]{.8} 14 \\
\hline
  {\bf 26} &          3 &          3 &    \cellcolor[gray]{.6} 2 &          3 &          - &          - &         \cellcolor[gray]{.8} 3 \\
\hline
  {\bf 27} &          2 &          2 &          4 &          - &          2 &    \cellcolor[gray]{.6} 2 &         \cellcolor[gray]{.8} 2 \\
\hline
  {\bf 28} &         72 &         83 &        139 &          - &        128 &   \cellcolor[gray]{.6} 72 &        \cellcolor[gray]{.8} 15 \\
\hline
    {\bf } &     {\bf } &     {\bf } &     {\bf } &     {\bf } &     {\bf } &     {\bf } &     {\bf } \\

    {\bf } &   {\bf W1} &     {\bf } &     {\bf } &   {\bf W4} &   {\bf W5} &     {\bf } &   {\bf W7} \\

{\bf Line-Station:} & {\bf Line1-s3} &     {\bf } &     {\bf } & {\bf Line1-s2} & {\bf Line1-s1} &     {\bf } & {\bf Line1-s4} \\
\hline
{\bf Cycle time:} &        135 &            &            &        135 &        124 &            &        104 \\

    {\bf } &     {\bf } &     {\bf } &     {\bf } &     {\bf } &     {\bf } &     {\bf } &     {\bf } \\

    {\bf } &     {\bf } &   {\bf W2} &   {\bf W3} &     {\bf } &     {\bf } &   {\bf W6} &     {\bf } \\

{\bf Line-Station:} &     {\bf } & {\bf Line2-s2} & {\bf Line2-s1} &     {\bf } &     {\bf } & {\bf Line2-s3} &     {\bf } \\
\hline
{\bf Cycle time:} &            &        354 &        347 &            &            &        354 &            \\
\hline
\end{tabular}
}
\caption{Matrix of process times for HESKIA\_64 and solution with two parallel lines}
\label{tab:heskia_64}
\end{table}

In the table we have represented an alternative solution by using a light shade on cells with task-assignments within line 1, and a darker shade on those cells with task-assignments within line 2; and where it can be easily checked how precedence constraints have been respected in both parallel lines. At the bottom of the table we summarized the workstations of each line with the following independent work teams:
\begin{itemize}
	\item Line 1 is composed of four consecutive stations with workers W5, W4, W1 and W7. Workers W1 and W4 are bottlenecks with a cycle time of 135.
	\item Line 2 is composed of three consecutive stations with workers W3, W2, and W6. Workers W2 and W6 are bottlenecks with a cycle time of 354.
\end{itemize}

To compute the combined cycle time (CT) of this configuration we can define the corresponding Throughput Rates ($TR$) of both lines. In this case, we have: 
\begin{itemize}
	\item Line 1 has $CT_1$ = 135 s/product, which is equivalent to a throughput  $TR_1$ = 26.67 products/h
	\item Line 2 has $CT_2$ = 354 s/product, which is equivalent to a throughput $TR_2$ = 10.17 products/h
\end{itemize}

Thus, the combined Throughput Rate using this parallelized configuration is:
$TR = TR_1 + TR_2$ = 36.84 products/h, which corresponds to a combined cycle time $CT$ = 97.7 s/product.

Therefore, the optimal cycle time of 126 obtained for HESKIA\_64 with the traditional serial ALWABP approach is improved by more than 22\% just by dividing the available workers in two lines. As will be later demonstrated, this is not an isolated successful example, since many others benchmark cases get better cycle times through this strategy.

\subsection{Review on parallelization within assembly lines}

Despite the fact that most assembly lines addressed in the literature are quite different from ALWABP, it is important to review those references that face parallelization and are somehow close to our approach. Thus, we can start chronologically by citing the pioneer work of \cite{buzacott1990abandoning}, that was the first to present models that incorporate not only parallel stations but asynchronous workflow and small inventories; also analyzing the problems in synchronizing parts delivery and the impact of job resequencing requirements. \cite{daganzo1994assembly} develop an analytical model to evaluate serial and parallel configurations, establishing some principles for assembly system design based on a trade-off between labour and equipment costs. However, as it has been introduced, the cost of human resources in SWDs scenario includes social dimensions not easy to quantify, what makes this analytical model unsuitable. Also \cite{bukchin03weighted} study the equipment selection problem on parallel workstations, investigating the influence of assembly sequence flexibility and cycle time on the balancing improvement due to the station paralleling. \cite{boysen08versatile} present a versatile graphic algorithm that can be adapted for parallel stations, and \cite{ege09assembly} consider parallel workstations and propose a branch and bound algorithm again based on the minimization of total equipment and workstation opening costs. \cite{kaku2008study} analyse the human-task-related performances in converting conveyor assembly lines to parallel lines. \cite{lusa2008survey} presents a survey on the multiple or parallel assembly lines balancing problem. As the author points out, the literature for these problems is quite poor, due to their difficulty.

Even excluding from this review the stochastic concerns and most mixed model focused approaches (as we have homogeneous product to be performed in parallel assembly lines) the Parallel Assembly Line Balancing Problem (PALBP), where different models can be assigned to parallel lines while considering a joint cycle time, should be mentioned as the main topic of research on parallelization in the last years (e.g. \cite{gokcen06balancing}; \cite{scholl09designing}; \cite{kara2010balancing}; and \cite{ozbakir11multiple}).

To the best of our knowledge, the only reference that studies the ALWABP with parallelization is \citep{araujo12two}, where two extensions are explored: in the first extension the authors allow multiple stations to execute the same set of tasks at a given point in the line; and in the second extension different workers can complement their capabilities collaborating in the same product within one workstation. The main difference of their first extension with respect to our proposal is that we force to configure complete parallel assembly lines, building up independent work teams that do not increase complexity of control. Note that solving this new problem requires very different techniques, since one of the most important decisions now is the division of the workers among the lines, which was not necessary in the problem tackled by \cite{araujo12two}. 

In the following section, we present a formal definition of this problem by means of a classical literature taxonomy and also via a mathematical model.

\section{Formal definition and mathematical model} \label{problem}

We formally classify the PALWABP using the nomenclature of \cite{boysen07classification}, that structures the vast field of assembly line balancing problems by means of a notation consisting of three elements [$\alpha$$\vert$$\beta$$\vert$$\gamma$], where: 
\begin{itemize}
	\item $\alpha$ concerns the precedence graph characteristics;
	\item $\beta$ concerns the station and line characteristics;
	\item and $\gamma$ concerns the optimization objectives.
\end{itemize}

\cite{boysen07classification} classified the ALWABP-2 as [pa,link,cum$\vert$equip$\vert$c]. In the PALWABP, we allow multiple assembly lines ($\beta$=pline). Therefore, the PALWABP-2 can be classified as [pa,link,cum$\vert$pline,equip$\vert$c]; whereas the previous proposal of \cite{araujo12two} has formal and practical differences that have been properly detailed, being coded as [pa,link,cum$\vert$ pstat,equip$\vert$c].

\subsection{Mathematical model for PALWABP}

Let $(N,\leq)$ be a partially ordered set of tasks in which $i< (>) j$ indicates that a task i must precede (succeed) a task j.  Also let $W$ be a set of workers and $p_{wi}$ an integer associated with each pair $(w,i) \in (W,N)$ indicating the time worker $w$ spends to execute task $i$. Also, $I_w$ is the set of tasks that worker $w$ is not able to perform. Consider an assignment  $a_t:N\rightarrow S$ of the tasks to a linear sequence of stations $S=\{1,2,\ldots,m\}$ respecting the tasks partial order and an assignment $a_w:W\rightarrow S$ of the workers to the same linear sequence of stations $S=\{1,2,\ldots,m\}$. The load of a station is the time the worker assigned to that station needs to execute those tasks assigned to the same station. Finally, the \emph{cycle time} of such  assignments is the largest load among all stations. The ALWABP-2 aims at finding assignments $a_t$ and $a_w$ minimizing the line cycle time. Let $\textrm{ALWABP-2}(N,W,p_{wi})$ be such optimal cycle time.
 
The PALWABP-2 considers the same input data and an extra integer, $K_{max}$, which limits the maximum number of parallel assembly lines. It then aims at finding a partition of $W$ in  $W_1, W_2... W_{K_{max}}$ sets, such that the overall throughput rate $\sum_{k=1}^{K_{max}} \frac{1}{\textrm{ALWABP-2}(N,W_k, p_{wi})}$ is maximal. We assume that a set $W_k$ can be empty and, in this case, $\textrm{ALWABP-2}(N,W_k, p_{wi}) = \infty$. In practice, $K_{max}$ is associated to the practical aspects of implementing multiple assembly lines such as the cost of tools or the available physical space.

In the following, we present a mathematical model for the PALWABP-2.  We define the following set of variables:

\vspace{0.2cm}

\begin{tabular}[t]{lp{10cm}}
$x_{swik}$ & Binary variable. Equals one if worker \textit{w} executes task \textit{i} in station \textit{s} of line \textit{k} and zero otherwise \\
$y_{swk}$ & Binary variable. Equals one if worker \textit{w} is designated to station \textit{s} in line \textit{k} and zero otherwise\\
$z_k$ & Binary variable. Equals one if line \textit{k} is active (that is, if at least one worker is assigned to that line) and zero otherwise \\
$C_k$ & The cycle time of line \textit{k}\\
\end{tabular}

\vspace{0.2cm}

We can write the following model for the ALWABP with parallel lines:

\begin{align}
& \displaystyle Min \frac{1}{\sum_{k = 1 | z_k > 0}^{K_{max}}\frac{1}{C_k}}\label{eq5:objetivo0}
\end{align}

Subject to:

\begin{align}
& \displaystyle\sum_{w \in W} \sum_{s \in S} x_{swik} = z_k, & \forall i \in N, \forall k \in K, \label{eq5:garante_execucao}\\
& \displaystyle\sum_{s \in S} \sum_{k = 1}^{K_{max}} y_{swk} = 1, & \forall w \in W, \label{eq5:biunivoco1} \\
& \displaystyle\sum_{w \in W} y_{swk} \leq z_k, & \forall s \in S, \forall k \in K, \label{eq5:biunivoco2} \\
& \displaystyle\sum_{w \in W} \sum_{s \in S | s \geq t} x_{swik} \leq \sum_{w \in W} \sum_{s \in S | s \geq t} x_{swjk}, & \forall i,j \in N | i<j,\label{eq5:precedencia}\\
&\qquad &\qquad \forall t \in S, \forall k \in K, \nonumber\\
& \displaystyle\sum_{i \in N} x_{swik} \leq |N| y_{swk}, & \forall w \in W, \forall s \in S, \forall k \in K, \label{eq5:relacao_trabalhador_estagio}\\
& \displaystyle \sum_{i \in N} p_{wi}x_{swik} \leq C_{k}, & \forall s \in S, \forall w \in W, \forall k \in K, \label{eq5:define_frequencia}\\
& \displaystyle x_{swik} = 0, & \forall w \in W, \forall s \in S,\label{eq5:incompativeis}\\
&\qquad &\qquad \forall i \in I_w, \forall k \in K, \nonumber\\
& \displaystyle x_{swik} \in \{0, 1\}, & \forall s \in S, \forall w \in W, \label{eq5:var1}\\
&\qquad &\qquad \forall i \in N, \forall k \in K, \nonumber\\
& \displaystyle y_{swk} \in \{0, 1\}, & \forall s \in S, \forall w \in W, \forall k \in K, \label{eq5:var2}\\
& \displaystyle z_k \in \{0, 1\}, & \forall k \in K. \label{eq5:var3}
\end{align}

The objective function (\ref{eq5:objetivo0}) minimizes the combined cycle time. Constraints (\ref{eq5:garante_execucao}) guarantee that in every active line, each task is executed by a single worker in a single station. Constraints (\ref{eq5:biunivoco1}) guarantee that every worker will be assigned to only one station from one of the parallel lines. Constraints (\ref{eq5:biunivoco2}) allow a single worker in each station of active lines and zero in each station of non-active lines. Constraints (\ref{eq5:precedencia}) establish the precedence relations between the tasks in each assembly line. These constrains were proposed by \cite{Ritt2011} and found out to be the most efficient among various linear precedence constraints alternatives. Constraints (\ref{eq5:relacao_trabalhador_estagio}) state that a worker can execute tasks in one stage only if that worker is assigned to that stage.  Constraints (\ref{eq5:define_frequencia}) define the cycle time for each assembly line as the maximum execution time among all workers from that line while constraints (\ref{eq5:incompativeis}) handle the task-worker incompatibilities.

The objective function (\ref{eq5:objetivo0}) makes this model non-linear. In order to linearize it, we adapted the linearization proposed by \cite{araujo12two} for the ALWABP with parallel workstations. In that work, the authors changed the objective function to maximize the throughput rate of the assembly line, which is equivalent to minimizing the cycle time. This can be adapted for the PALWABP by changing the objective function to maximize the sum of the throughput rates of each assembly line. Using the new variables:

\begin{tabular}[t]{lp{10cm}}
$F_{k}$ & The production rate of line \textit{k}.\\
$f_{swk}$ & The production rate of worker \textit{w} in stage \textit{s} of line \textit{k}. Equals zero if the worker is not assigned to that stage of that line.\\
$v_{swik}$ & Auxiliary variable used for linearization. Corresponds to $f_{swk}*x_{xwik}$.\\ 
\end{tabular}

Let $M$ be an upper bound for the production rate. A simple upper bound corresponds to:

$M = \sum_{i \in N} min(w)p_{wi}/|W|$

We can write a linear model for the PALWABP as:

\begin{align}
& \displaystyle Max \sum_{p = 1}^{K_{max}} F_k \label{eq5:objetivo}
\end{align}

Subject to:

\begin{align}
& (\ref{eq5:garante_execucao}) - (\ref{eq5:var3})&\\
& \displaystyle F_k \leq M z_k & \forall k \in K \label{eq5:define_frequencia2}\\
& \displaystyle \sum_{w \in W} f_{swk} \geq F_k - M(1 - \sum_{w \in W} y_{swk}), & \forall s \in S, \forall k \in K, \label{eq5:define_frequencia2a} \\
& \displaystyle \sum_{i \in N} p_{wi}v_{swik} = y_{swk}, & \forall w \in W, \forall s \in S, \forall k \in K, \label{eq5:define_frequencia2b} \\
& \displaystyle f_{swk} \leq My_{swk}, & \forall w \in W, \forall s \in S, \forall k \in K, \label{eq5:define_frequencia2c} \\
& \displaystyle v_{swik} \geq f_{swk} - M(1 - x_{swik}), & \forall s \in S, \forall w \in W,\label{eq5:define_frequencia2d}\\
&\qquad &\qquad \forall i \in N, \forall k \in K, \nonumber \\
& \displaystyle v_{swik} \leq Mx_{swik}, & \forall s \in S, \forall w \in W,\label{eq5:define_frequencia2e}\\
&\qquad &\qquad \forall i \in N, \forall k \in K, \nonumber\\
& \displaystyle F_k \geq 0, & \forall k \in K,\label{eq5:var4}\\
& \displaystyle f_{swk} \geq 0, & \forall s \in S, \forall w \in W, \forall k \in K,\label{eq5:var5}\\
& \displaystyle v_{swik} \geq 0, & \forall s \in S, \forall w \in W,\label{eq5:var6}\\
&\qquad &\qquad \forall i \in N, \forall k \in K. \nonumber
\end{align}

The new objective function (\ref{eq5:objetivo}) maximizes the overall production rate. Constraints \ref{eq5:define_frequencia2} ensure that the production rate of an empty line is zero. Constraints (\ref{eq5:define_frequencia2a}) define the production rate of active line as the production rate of the slowest stage. Constraints (\ref{eq5:define_frequencia2b}) define $v_{swik}$ weighted by the task times. Constraints (\ref{eq5:define_frequencia2c}) state that $f_{swk} = 0$ if worker \textit{w} is not assigned to stage \textit{s} of line \textit{k}. Constraints (\ref{eq5:define_frequencia2d}) and (\ref{eq5:define_frequencia2e}) set the bounds of $v_{swik}$. Constraints (\ref{eq5:define_frequencia2c}) and (\ref{eq5:define_frequencia2d}) make $v_{swik} = f_{swk}$ if $x_{swik} = 1$, while constraints (\ref{eq5:define_frequencia2e}) make $v_{swik} = 0$ if $x_{swik} = 0$.

This model is used to solve small instances (see Section~\ref{computational}). Nevertheless, due to its complexity, heuristic methods were developed in order to solve instances of practical size. These methods are described in the following two sections.

\section{Heuristic methodologies}\label{methodologies}

In this section, we describe a tabu search method and a biased random-key genetic algorithm for the PALWABP.

\subsection{The tabu search algorithm}\label{heuristics}

In this section we present a tabu search method to solve the PALWABP-2. The core idea of the method is the generation of sets of workers that can be assigned together in an independent line. Each of such team of workers must be able to execute all tasks while respecting the precedence constraints.

In this method there are two preprocessing steps. In the first step we generate sets of tasks that each worker can execute. In the second step, we use these sets of tasks to generate feasible sets of workers (i.e., workers that can be assigned together to a complete line) with minimum cardinality. The following sections detail the preprocessing steps and the main method.

\subsubsection{Feasible worker-tasks sets generation}

In this first step, the sets of all tasks that can be executed in a single station by each worker are obtained. If cycle time constraints are ignored, a worker $w$ can execute any set of tasks as long as he is not obliged to execute a task in $I_w$. Let ${i\in N}$ be a task and $I \subset N$ be a set of tasks.  We use the notation $i \not> (\not<) I$ to indicate that task $i$ does not succeed (precede) any of the tasks in $I$ and the notation $i >> s$ to indicate that task $i$ is an immediate successor of a task in $s$.  The method presented in Algorithm~\ref{alg:task_sets_generation} is able to determine $T_w$, the sets of tasks that can be executed by a worker $w$.

\begin{algorithm}[!ht]
\caption{Task sets generation}
\label{alg:task_sets_generation}
\begin{algorithmic}[1]
\REQUIRE $ N, w \in W$ 
\STATE $T_w = \emptyset $
\STATE $I = \emptyset$
\REPEAT
		\STATE $s = \{i \in N\backslash I_w | i \not< I, i \not> I_w\backslash I\}$\label{step:tarefas_disponiveis}
		\STATE $T_w = T_w \cup s$\label{step:adiciona_set}
		\STATE $I = \{i \in I_w | i >> s\}$\label{step:prox_subset}
\UNTIL{$I = \emptyset$}
\STATE Return $T_w$
\end{algorithmic}
\end{algorithm}

The algorithm receives as input the worker being analyzed and the ordered set of tasks. It considers subsets of $I_w$ in order to generate sets of tasks that worker $w$ can execute. For each subset $I$ considered, the algorithm assumes that the tasks in $I$ have been executed in stations preceding the station to which $w$ has been assigned and calculates $s$, the set of tasks that $w$ can execute which contains the tasks that do not succeed any other task that $w$ can not execute (line \ref{step:tarefas_disponiveis}). This set is then added to the list of sets (line \ref{step:adiciona_set}). The next step of the algorithm is to generate the next subset $I$ based on the set of tasks generated in this iteration. The new subset $I$ contains all infeasible tasks that are immediate successors of at least one task belonging to the current set $s$ (line \ref{step:prox_subset}). This process is repeated until $I = \emptyset$, which means no infeasible tasks succeed the tasks in the current set $s$.

While this algorithm may not generate all possible sets of tasks we found that the ones it generates are enough to feed the next steps of the method. It also helps the next preprocessing step to quickly identify instances in which there is no feasible solutions with more than one parallel assembly line. The following example illustrates this method.

\begin{figure}[!ht]
\centering
\resizebox{\textwidth}{!}{
\includegraphics{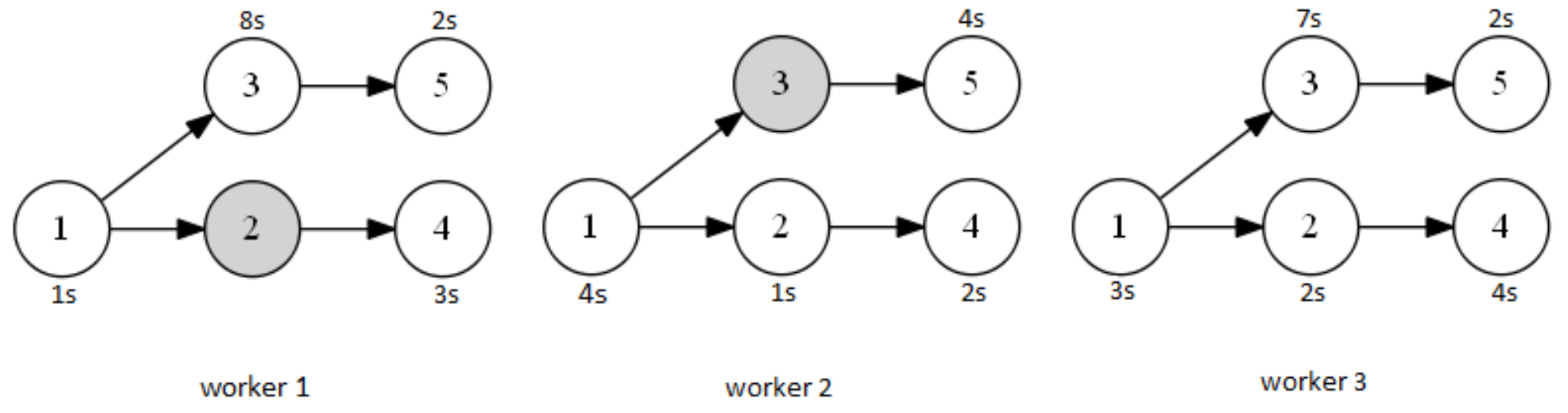}
}
\caption{Example of precedence graph}
\label{fig:ex_grafo}
\end{figure}

\paragraph{Example:} Consider the precedence graph shown in Figure \ref{fig:ex_grafo}. Worker $w_1$  can not execute task $2$ while worker $w_2$ can not execute task $3$.

The subsets of $I_{w_1}$ and $I_{w_2}$ (see line 2 of Algorithm~\ref{alg:task_sets_generation})  are $\{\emptyset, \{2\}\}$ and $\{\emptyset, \{3\}\}$, respectively. Task $\emptyset$ indicating an artificial task, is added to these sets to represent the case in which the worker is assigned to the first station in the assembly line. They yield $T_{w_1} = \{\{\emptyset,1,3,5\},\{3,4,5\}\}$ and  $T_{w_2} = \{\{\emptyset,1,2,4\},\{2,4,5\}\}$. Worker $w_3$ can execute all tasks and, therefore, $T_{w_3} = \{\{\emptyset,1,2,3,4,5\}\}$

\subsubsection{Feasible worker sets generation}

The goal of this step is to select one set of tasks from each worker so that each task is covered by at least one set. By doing this, we want to ensure the existence of a viable solution for the ALWABP with these workers. The information obtained in the previous stage is now used to determine all combinations of workers that can be assigned together to a team. We propose Algorithm \ref{alg:gera_subproblemas} to generate such sets.

\begin{algorithm}[!ht]
\caption{Subproblems generation}
\label{alg:gera_subproblemas}
\begin{algorithmic}[1]
\REQUIRE $T_w$
\STATE $W_r = \emptyset$, $set\_list = \emptyset$, $worker\_list = \emptyset$
\STATE $i = 0$\label{step:artificial_task}
\FORALL{$w \in W - worker\_list$}
	\FORALL{$s \in T_w | i \in s$}\label{step:select_set}
		\STATE $set\_list = set\_list \cup \{s\}$ \label{step:add_set}
		\STATE $worker\_list = worker\_list \cup \{w\}$ \label{step:add_worker}
		\IF{$((N - set\_list) = \emptyset)$}
			\STATE $W_r = W_r \cup \{worker\_list\}$ \label{step:add_subproblem}
			\STATE Make w = the last element from $worker\_list$ and s = the last element from $set\_list$
			\STATE Remove last element from $worker\_list$ and $set\_list$\label{step:remove_worker}
		\ELSE
			\STATE Select $i \in (N - set\_list)$\label{step:select_task}
		\ENDIF
	\ENDFOR
\ENDFOR
\IF{$set\_list \not= \emptyset$}
	\STATE Make w = the last element from $worker\_list$ and s = the last element from $set\_list$
	\STATE Remove last element from $worker\_list$ and $set\_list$ \label{step:remove_last}
	\STATE goto \ref{step:select_set}
\ELSE
	\STATE Return $W_r$\label{step:gera_subp_fim}
\ENDIF
\end{algorithmic}
\end{algorithm}

This algorithm enumerates all possible combinations of workers that may be assigned to a team. We start with the artificial task (line \ref{step:artificial_task}) as the next task to be considered. In each iteration we select a set from the available workers that contains the current task (line \ref{step:select_set}). The set and the worker are added to their respective lists (lines \ref{step:add_set} and \ref{step:add_worker}). If the sets in $set\_list$ cover all tasks, the current list of workers can generate a feasible solution and it is added to the list of subproblems to be solved in the next step (line \ref{step:add_subproblem}). Otherwise, the method continues with a new uncovered task (line \ref{step:select_task}). If no available set covers the current task, the algorithm removes the last worker and set in their respective lists (line \ref{step:remove_last}). Then, the algorithm return to line \ref{step:select_set} and selects another set. Observe that, when the algorithm returns to line \ref{step:select_set}, it selects another set. This process continues until the algorithm generates all combinations of workers (line \ref{step:gera_subp_fim}).

\newpage
\paragraph{Example (continued):} For the example of Figure \ref{fig:ex_grafo}, all possible obtained combinations of sets are:

\begin{center}
\begin{tabular}{c|ccc}
Line & worker $w_1$ & worker $w_2$ & worker $w_3$ \\ \hline
1 & $T_{w_1}(1) = \{\emptyset,1,3,5\}$  & $T_{w_2}(1) = \{\emptyset,1,2,4\}$ & \\
2 & $T_{w_1}(1) = \{\emptyset,1,3,5\}$  & $T_{w_2}(2) = \{2,4,5\}$ & \\
3 & $T_{w_1}(2) = \{3,4,5\}$  & $T_{w_2}(1) = \{\emptyset,1,2,4\}$ & \\
4 & $T_{w_1}(1) = \{\emptyset,1,3,5\}$  &  &  $T_{w_3}(1) = \{\emptyset,1,2,3,4,5\}$\\
5 & $T_{w_1}(2) = \{3,4,5\}$  &  &  $T_{w_3}(1) = \{\emptyset,1,2,3,4,5\}$  \\
6 &   & $T_{w_2}(1) = \{\emptyset,1,2,4\}$ &  $T_{w_3}(1) = \{\emptyset,1,2,3,4,5\}$\\
7 &   &  $T_{w_2}(2) = \{2,4,5\}$   &  $T_{w_3}(1) = \{\emptyset,1,2,3,4,5\}$  \\
8 & $T_{w_1}(1) = \{\emptyset,1,3,5\}$  & $T_{w_2}(1) = \{\emptyset,1,2,4\}$ & $T_{w_3}(1) = \{\emptyset,1,2,3,4,5\}$\\
9 & $T_{w_1}(1) = \{\emptyset,1,3,5\}$  & $T_{w_2}(2) = \{2,4,5\}$ & $T_{w_3}(1) = \{\emptyset,1,2,3,4,5\}$\\
10 & $T_{w_1}(2) = \{3,4,5\}$  & $T_{w_2}(1) = \{\emptyset,1,2,4\}$ & $T_{w_3}(1) = \{\emptyset,1,2,3,4,5\}$ \\
11 & $T_{w_1}(2) = \{3,4,5\}$  & $T_{w_2}(1) = \{2,4,5\}$ & $T_{w_3}(1) = \{\emptyset,1,2,3,4,5\}$ \\
12 &    &   & $T_{w_3}(1) = \{\emptyset,1,2,3,4,5\}$
\end{tabular}
\end{center}

Each of these combinations result in at least one feasible ALWABP solution.

\subsubsection{Solution evaluation and neighborhood structure}

The tabu search method developed explores the space defined by the workers partition. Each solution defines the line  to which each worker is assigned  and is evaluated with a simple constructive heuristic developed by \cite{moreira12hybrid} for the serial ALWABP. This method generates feasible solutions by sequentially assigning workers and tasks to the stations in a order defined by some heuristic criteria. At each iteration a tentative cycle time is used (starting from a know lower bound) and the method increases this tentative value by one unity if a feasible solution is not found. In our method, this heuristic is modified so that at each unsuccessful iteration the tentative cycle time is increased by the minimum available task time, i.e., $\bar{c} = \bar{c} + \min_{w,i}{t_{wi}}$, which greatly improves the method speed with little effect on solution quality.

To generate a feasible initial solution for the PALWABP we must select $K_{max}$ solutions for the ALWABP, which are partial solutions for the PALWABP. Algorithm \ref{alg:gera_solucao_inicial} selects which sets of workers generated in the previous step will form the base of a feasible solution.

\begin{algorithm}[!ht]
\caption{Initial solution for the tabu search}
\label{alg:gera_solucao_inicial}
\begin{algorithmic}[1]
\REQUIRE $W_r$, $K_{max}$
\STATE $list = \emptyset$
\STATE Select $W_i$ the first element in $W_r$
\WHILE{$|list| < K_{max}$}
	\IF{$W_i \cap W_j = \emptyset$ $\forall W_j \in list$}\label{step:verifica_sobrepoe}
		\STATE $list = list \cup {W_i}$
	\ENDIF
	\IF{$|list|+1 = K_{max}$}
		\STATE Return $list$\label{step:lista_completa}
	\ELSE
		\IF{$W_i$ is not the last element of $W_r$}
			\STATE Select $W_i$ the next element of $W_r$
		\ELSE 
			\IF{$list \neq \emptyset$}
				\STATE Make $W_i$ the element in $W_r$ after the last element in $list$\label{step:seleciona_prox}
				\STATE Remove the last element in $list$\label{step:remove_ultimo}
			\ELSE 
				\STATE ERROR: No solution exists\label{step:sem_solucao}
			\ENDIF
		\ENDIF
	\ENDIF
\ENDWHILE
\end{algorithmic}
\end{algorithm}

This algorithm starts with a empty list of new sets. It then adds the first element in $W_r$ that does not overlap with any sets currently in the list (step \ref{step:verifica_sobrepoe}). This process continues until we have a list of $K_{max}$ sets (line \ref{step:lista_completa}). If the algorithm reaches the end of $W_r$ without finding a list of $K_{max}$ sets it removes the last element in the list and tries to continue with the next element (lines \ref{step:seleciona_prox} and \ref{step:remove_ultimo}) or finds that no solution exists (line \ref{step:sem_solucao}). Any workers that are not assigned to a set in the list are assigned randomly. We then use the modified constructive heuristic to generate feasible partial solutions, which form a feasible initial solution for the PALWABP.

A neighborhood solution is obtained with two movements: 1) moving a worker from one assembly line to another and 2) swapping two workers between lines. Algorithm \ref{alg:vizinhanca1} calculates the neighborhood of a given solution.

\begin{algorithm}[!ht]
\caption{Neighborhood}
\label{alg:vizinhanca1}
\begin{algorithmic}[1]
\REQUIRE $W_r$
\FORALL{$W_{i1} \in solution$}
	\FORALL{$W_j \in W_r | W_{i1} \cap W_j \not=\emptyset$}
		\STATE $excess = W_{i1}-W_j$, $lack = W_j-W_{i1}$
		\IF{$excess\not=\emptyset$}\label{step:verifica_diferente}
			\IF{$lack=\emptyset$}
				\STATE Add all workers in $excess$ to $neighborhood$\label{step:movimento_simples}
			\ELSE
				\FORALL{$W_{i2} \in solution | W_{i2} \not= W_{i1}$}
					\IF{$lack \subset W_{i2}$ \AND $((W_{i2} - lack) \bigcup excess)$ is feasible}
						\STATE Add to $neighborhood$ the movement: move the workers in $lack$ to $i1$ and move the workers in $excess$ to $i2$\label{step:movimento_troca}
					\ENDIF
				\ENDFOR
			\ENDIF
		\ENDIF
	\ENDFOR
\ENDFOR
\STATE Return $neighborhood$
\end{algorithmic}
\end{algorithm}

Each set of workers generated in the preprocessing step contains the minimum amount of workers to execute all tasks in a independent assembly line. Therefore, by comparing the sets of workers generated in the preprocessing step with the sets of workers in the current solution (step \ref{step:verifica_diferente}), we can verify which workers can be moved between the assembly lines. While some of the generated sets may result in unfeasible solutions, this algorithm boosts the performance of the tabu search by excluding most unfeasible movements.

The method then selects the non-tabu movement that lead to the best solution in the neighborhood. However, we may select a tabu movement if it leads to a better solution than the incumbent solution. The movement is then added to tabu list for a given number of iterations. In the next section, we present another heuristic developed for the PALWABP.

\subsection{The biased random-key genetic algorithm (BRKGA)}\label{brkga}

In this section, we present a BRKGA for the PALWABP. This metaheuristic is a variation of the genetic algorithm in which we use a string of randomly generated numbers between 0 and 1 to represent each individual. The population is divided in elite and non-elite individuals and the crossover always happen between individuals of different categories. One advantage of this method is that we only need to provide the fitness function and the method parameters (size of the population, number of elite individuals, chance of mutation and chance of inheriting an allele from elite parent).

For the fitness function we must first determine how to translate the string of numbers in each chromosome into a feasible solution. The main decisions in PALWABP consist on selecting the workers to be assigned to each parallel line and determining the configuration of such lines. In order to do that, we split the chromosome in three parts. The first part has $|W|$ genes and defines which workers are assigned to each parallel line. The gene $c[w]$ determines that we must assign worker $w$ to line $\lfloor c[w] * K_{max} \rfloor$.

The second and third part of the chromosome are used to solve the subproblem in each line. They replace the heuristic criteria used in the constructive heuristic by \cite{moreira12simple}. The second part of the chromosome has $|W|$ genes and defines the priorities of the workers in each assembly line. Observe that since each worker must be assigned to exactly one line we do not need to duplicate these data. The third part of chromosome has $|N| * K_{max}$ genes and defines the task priorities for each line.

\section{Computational results}\label{computational}

In this section we present the results obtained using the mathematical model and the heuristics presented in sections \ref{problem} and \ref{methodologies}, respectively. To test these strategies, we generated a set of instances for the ALWABP based on the instances provided by \cite{otto13systematic} for the SALBP. These instances vary in graph structure (\textit{bottleneck}, \textit{chain} or \textit{mixed}), `trickiness', order of strength and times distribution; representing a much more robust SALBP benchmark than the classical benchmark of \cite{hoffmann90assembly}, as demonstrated by the authors. For each instance, we generated four instances for the ALWABP, with different time factors (2 or 5) and infeasibility rates (10\% or 20\%). A time factor of \textit{n} means that, for every task with execution time of \textit{t} in the original instance, the execution time of each worker is a random integer between \textit{t} and \textit{n * t}. The number of workers in each instance is equal to the number of stations in the best known solution for the original instance for the SALBP-1 with a cycle time of 1000s. We used the sets of instances with 20 tasks and 0.2 order strength as our base for generating a total of 900 instances.

The model was run using IBM ILOG Cplex 12.5. Both heuristic strategies were implemented in C++ and tested in a computer with a Intel Core 2 Duo T5450 processor, 1,66 GHz and 3 GB of RAM. The following sections detail the obtained results. These are compared with the best know solutions for the serial problem.

\subsection{Results for the mathematical model}

The model presented in section 3 was implemented and used to solve the 225 easier instances generated with 10\% infeasibility rate and a time factor of 2. The method was given a time limit of 1800 seconds for each instance.

The model proved to be very difficult to solve. After 30 minutes of execution, it could prove optimality for only 20 instances. It also managed to prove that there was no feasible solution using more than one assembly line for 73 instances. For the remaining instances we compared the cycle time of the best solution found with the lower bound (that is, the cycle time of the best relaxed solution). We found that the gap between these values was very high, usually above 1000\%. This means that Cplex could not tell how close these solutions were from the optimal solution. The model was unable to find solutions whose cycle time was lower than the best known cycle time for the respective serial ALWABP.

\subsection{Results for the heuristics}

For the tabu search we randomly select a set of workers in Algorithm \ref{alg:gera_solucao_inicial}. This set is added to the list of sets and cannot be removed. We then run the method for as many iterations as necessary, until 1000 iterations have passed without changing the incumbent solution. The method then restarts with a new random solution, which happens up to 10 times. We used a tabu list of 10 movements. For the BRKGA, we used a population of 100 individuals and an elite population of 20 individuals. Ten individuals are replaced by mutants at each iteration. Each gene of a new individual was copied from the elite parent with a 70\% probability.

Tables \ref{tab:results_tabu} and \ref{tab:results_brkga} present the results obtained by the tabu search method and the BRKGA described in Sections \ref{heuristics} and \ref{brkga}, respectively, for the whole set of instances, using two parallel lines. As a comparison benchmark, table \ref{tab:results_enum} also presents the results obtained through brute force enumeration of all combinations of subproblems generated by the preprocessing steps of the tabu search method. As for the other methods, the subproblems were solved using the constructive heuristic developed by \cite{moreira12simple}. The results are grouped by task times variation, percentage of incompatible tasks and type of precedence graph. For the tabu search method, each instance was solved 5 times and the average results are presented. Columns C\% represent the average difference between the best solution found by our method in each execution and the best known solution for the serial ALWABP. Columns T indicate the average execution time of the five executions in seconds. Columns P\% indicate the percentage of solutions using parallel lines found, that is, the percentage of instances solved by the method. Column Best C\% represent the average difference between the best solution found by our method among the five executions and the best known solution for the serial ALWABP and column SD represents the average standard deviation.

\begin{table}[!htbp]
  \centering
  \caption{Results for the tabu search method}
    \begin{tabular}{|r|r|r|r|r|r|r|r|}
\hline
          &       &       & C\%   & T     & P\%   & Best  & SD \\
\hline
    t-2t  & 10\%  & Bottleneck & 1.53\% & 88    & 59.46\% & -13.76\% & 0.81\% \\
          &       & Chain & 3.53\% & 67    & 68.92\% & -14.90\% & 0.94\% \\
          &       & Mixed & 2.20\% & 81    & 64.86\% & -13.86\% & 0.90\% \\
          & 20\%  & Bottleneck & -1.35\% & 106   & 40.54\% & -15.09\% & 0.84\% \\
          &       & Chain & 1.94\% & 80    & 45.95\% & -15.64\% & 0.86\% \\
          &       & Mixed & 0.08\% & 97    & 45.95\% & -11.35\% & 0.95\% \\
\hline
    t-5t  & 10\%  & Bottleneck & 9.53\% & 67    & 59.46\% & -20.95\% & 0.70\% \\
          &       & Chain & 11.34\% & 44    & 68.92\% & -15.40\% & 0.98\% \\
          &       & Mixed & 11.64\% & 55    & 64.86\% & -10.33\% & 0.81\% \\
          & 20\%  & Bottleneck & 6.69\% & 87    & 40.54\% & -11.29\% & 1.54\% \\
          &       & Chain & 9.13\% & 64    & 45.95\% & -13.10\% & 0.70\% \\
          &       & Mixed & 8.60\% & 81    & 45.95\% & -10.11\% & 0.88\% \\
\hline
    \end{tabular}%
  \label{tab:results_tabu}%
\end{table}%
\begin{table}[!htbp]
  \centering
  \caption{Results for the BRKGA}
    \begin{tabular}{|r|r|r|r|r|r|r|}
\hline
          &       &       & C\%   & T     & P\%   & Best \\
\hline
    t-2t  & 10\%  & Bottleneck & 7.18\% & 231   & 59.46\% & -9.48\%  \\
          &       & Chain & 10.41\% & 277   & 68.92\% & -11.98\% \\
          &       & Mixed & 8.03\% & 263   & 64.86\% & -7.60\% \\
          & 20\%  & Bottleneck & 5.04\% & 146   & 40.54\% & -7.89\%  \\
          &       & Chain & 7.13\% & 196   & 45.95\% & -13.57\% \\
          &       & Mixed & 6.36\% & 191   & 45.95\% & -9.60\% \\
\hline
    t-5t  & 10\%  & Bottleneck & 20.22\% & 378   & 59.46\% & -7.62\% \\
          &       & Chain & 23.70\% & 498   & 68.92\% & -5.33\% \\
          &       & Mixed & 24.51\% & 463   & 64.86\% & -0.81\% \\
          & 20\%  & Bottleneck & 16.86\% & 238   & 40.54\% & -7.18\% \\
          &       & Chain & 20.02\% & 301   & 45.95\% & 0.31\% \\
          &       & Mixed & 19.89\% & 303   & 45.95\% & -0.17\% \\
\hline
    \end{tabular}%
  \label{tab:results_brkga}%
\end{table}%

\begin{table}[!htbp]
  \centering
  \caption{Results for the enumerative method}
    \begin{tabular}{|r|r|r|r|r|r|r|}
\hline
          &       &       & C\%   & T     & P\%   & Best \\
\hline
    t-2t  & 10\%  & Bottleneck & -0.25\% & 35    & 58.11\% & -15.37\%\\
          &       & Chain & 1.93\% & 24    & 68.92\% & -17.56\% \\
          &       & Mixed & 0.36\% & 35    & 64.86\% & -15.10\%  \\
          & 20\%  & Bottleneck & -2.02\% & 16    & 39.19\% & -14.94\% \\
          &       & Chain & 0.23\% & 15    & 45.95\% & -15.73\% \\
          &       & Mixed & -1.08\% & 21    & 45.95\% & -16.15\%  \\
\hline
    t-5t  & 10\%  & Bottleneck & 8.67\% & 45    & 58.11\% & -21.13\%\\
          &       & Chain & 10.09\% & 31    & 68.92\% & -15.85\% \\
          &       & Mixed & 11.19\% & 48    & 64.86\% & -10.67\%  \\
          & 20\%  & Bottleneck & 5.95\% & 21    & 39.19\% & -10.74\% \\
          &       & Chain & 8.64\% & 19    & 45.95\% & -13.10\% \\
          &       & Mixed & 8.27\% & 29    & 45.95\% & -9.96\%  \\
\hline
    \end{tabular}%
  \label{tab:results_enum}%
\end{table}%

By comparing columns P\% from the three tables, we can see that the tabu search method and the BRKGA managed to solve more \textit{bottleneck} instances than the enumerative method, since some of them caused this method to run out of memory. This means that, while the enumeration of all combinations of subproblems is feasible for small instances, it may become infeasible for larger instances.

The tabu search method was superior to the BRKGA in both execution time and solution quality. As expected, the more worker tasks incompatibilities, the harder it is to find solutions with parallel lines. However, in many instances with low infeasibility rates, the method found solutions with lower cycle times than the respective solution for the serial ALWABP. The tabu search found solutions with cycle time up to 20,95\% lower than the cycle time for the best known solution for the serial ALWABP against 13,57\% for the BRKGA. Instances with \textit{bottleneck} precedence graphs resulted in the lowest average difference between the solution found by the heuristic and the respective ALWABP best known solution. However, the method found a lower percentage of feasible solutions for this subset of instances when compared to \textit{chain} and \textit{mixed} instances. Instances with \textit{chain} precedence graphs resulted in the highest percentage of parallel feasible solutions.

These results show that, on average, instances with \textit{bottleneck} precedence graph resulted in the highest percentage of improved solutions. In contrast, instances with \textit{chain} type precedence graphs resulted in the lowest percentage of improved solutions. This means that, on average, the heuristics found solutions with better quality for the former and a high number of parallel solutions for the latter. In general, one can see that the strategy of having multiple parallel lines can have a positive effect in such cases. Averages gain vary from 2,67\% to 6,13\% with respect to the best serial solutions. These results indicate that such configurations should be considered when planning such assembly lines. It should be noted that with this approach, apart from the productivity increase, production managers gain many more potential assignments for job rotation purposes; what implicitly improves the workers' wellbeing.

\section{Conclusions}\label{conclusions}

We have defined, modeled and solved a new assembly line balancing problem named Parallel Assembly Line Worker Assignment and Balancing Problem. This problem is motivated by the context found in sheltered work centers for the disabled where workers with very different characteristics are assigned to assembly lines. The proposed model and algorithms indicate that the use of parallel assembly lines in such situation might improve productivity levels. These results encourage the study of other alternative layouts or the proposal of even more efficient algorithms, getting a more balanced workload of human resources involved. 

From a practical point of view, we provide an additional powerful approach for coping with workers heterogeneity while ensuring the highest productivity; what becomes crucial for many SWDs survival in the current economical context. These first results are also exportable to ordinary environments with heterogeneous workers, what is also foreseen as another interesting research line. Finally, the use of this approach to gain additional job rotation scenarios draws a promising further research line.

Initially, it would be interesting to improve the job rotation proposal of \cite{costa09job}, since the additional possible assignments obtained by these new approaches may help in finding better job rotation schedules.

\subsubsection*{Acknowledgments}

This research was supported by CAPES-Brazil and MEC-Spain (coordinated project CAPES DGU 258-12 / PHB2011-0012-PC) and by FAPESP-Brazil. The authors thank Dr. Marcus Ritt, from Universidade Federal do Rio Grande do Sul (UFRGS - Brazil), for providing the optimal solutions for the serial ALWABP. The authors also thank three anonymous reviewers for their comments which have helped improve this paper.

\bibliographystyle{dcu}
\bibliography{library2}

\end{document}